# Software Cost Estimation Framework for Service-Oriented Architecture Systems using Divide-and-Conquer Approach

Zheng Li
NICTA and UNSW
School of Computer Science and Engineering
Sydney, Australia
zheng.li@nicta.com.au

Jacky Keung
NICTA and UNSW
School of Computer Science and Engineering
Sydney, Australia
jacky.keung@nicta.com.au

*Abstract*—Due to the complexity of Service-Oriented Architecture (SOA), cost and effort estimation for SOA-based software development is more difficult than that for traditional software development. Unfortunately, there is a lack of published work about cost and effort estimation for SOA-based software. Existing cost estimation approaches are inadequate to address the complex service-oriented systems. This paper proposes a novel framework based on Divide-and-Conquer (D&C) for cost estimation for building SOA-based software. By dealing with separately development parts, the D&C framework can help organizations simplify and regulate SOA implementation cost estimation. Furthermore, both cost estimation modeling and software sizing work can be satisfied respectively by switching the corresponding metrics within this framework. Given the requirement of developing these metrics, this framework also defines the future research in four different directions according to the separate cost estimation sub-problems.

*Keywords-service-oriented architecture (SOA); software cost estimation; divide-and-conquer (D&C); framework*

## I. INTRODUCTION

Software cost estimation for Service-Oriented Architecture (SOA-based) software development confronts more challenges than for traditional software development. One of the main reasons is the architectural difference in SOA compared to traditional software development. Josuttis [8] has pointed out that distributed processing would be inevitably more complicated than non-distributed processing, and any form of loose coupling increases complexity. Meanwhile, the more complexity involved in a system, the more difficulty the designers or engineers have to understand the implementation process and thus the system itself [25]. In other words, people have to devote more effort to accurate manipulations when performing more complicated tasks. In practice, building a true heterogeneous SOA for a wide range of operating environments may take years of development time if the company does not have sufficient SOA experience and expertise [2]. What is more, it is difficult to foresee and justify the cost and effort of developing an SOA application before the project starts.

The problem of SOA cost estimation has not been addressed adequately in the existing literature. The current cost estimation approaches for traditional software development are inadequate for complex service-oriented software. For example, COCOMO II cannot arrive at global cost approximation for the entire SOA application development, and expert judgment may easily fall into traps of uncertainty or bias because of the complexity of the SOA.

This paper proposes a novel framework by employing a Divide-and-Conquer (D&C) method in an attempt to deal with cost estimation problem for SOA based software development. Within this D&C framework, services are classified into three primitive types and one combined type according to different development processes.

Cost estimation for developing primitive services can be handled as sub-problems that are small and independent enough to be solved. For combined services, the division procedure will occur recursively until all the resulting separated services are primitive. The cost and effort of service integration is then calculated gradually following the reverse division sequence.

In this paper, the application of the D&C cost estimation framework is demonstrated using a case study. The result shows that the proposed framework can simplify and regulate the complicated development cost estimation for SOA-based applications.

The next section reviews current approaches to software cost estimation for SOA-based software. Section III outlines the proposed framework. A case study is presented in section IV to demonstrate the application procedure proposed in the framework. Section V compares this D&C based solution with current cost estimation approaches for SOA applications. Section VI concludes and provides future research directions.

## II. RELATED WORK

### A. COCOMO II Related Approaches

COCOMO II (Constructive Cost Model) [4] is one of the best-known and best-documented algorithmic models, which allows organizations to estimate cost, effort, and schedule when planning new software development activities. Tansey and Stroulia [6] have attempted to use COCOMO II to estimate the cost of creating and migrating services. They reported that COCOMO II should be extended to accommodate new characteristics of SOA based development.

COCOMO II is generally inadequate to accommodate the cost estimation needs for SOA-based software development.

When considering the declarative composition specifications, a fundamentally different development process may be adopted in SOA-based software. Based on the Internet technologies, SOA-based software can be realized as a composition of loosely coupled services with well-defined interfaces and consistent communication protocols. These services hide technical details, and are not restricted to any specific technology. In other words, the service implementation is programming language and platform independent. Therefore, an SOA-based application could comprise the combination of all possible development strategies and development processes. Consequently, although the COCOMO II model has a large number of coefficients such as effort multipliers and scale factors, it is difficult to directly justify the cost estimation for SOA-based software development.

On the other hand, considering the difference between component orientation and service orientation [1], the COCOMO II model by itself is inadequate to estimate effort required when reusing service-oriented resources. COCOMO II considers two types of reused components, namely black-box components and white-box components. Black-box components can be reused without knowing the detailed code or making any change to it, while white-box components have to be modified with new code or integrated with other reused components before it can be reused. Similarly, within the SOA framework, there are black-box services that can be adopted directly, and white-box services that should be ported from legacy systems. Nevertheless, taking black-box reuse for instance, the difference between code-level and service-level reuse is significant. Whether a code-level component is suitable or not for reuse should be understood and revealed by using reverse engineering or re-engineering [5] according to the real situation. Comparatively, the contractually reusable and loosely coupled service can be reused directly through service discovery techniques, for example semantic annotation and quality of service.

### B. Function Point Analysis and Software Sizing

Size prediction for the constructed deliverables has been identified as one of the key elements in any software project estimation. SLOC (Source Line of Code) and Function Point are the two predominant sizing measures. Function Point measures software system size through quantifying the amount of functionality provided to the user in terms of the number of inputs, outputs, inquires, and files. In practice, Function Point can be used continuously throughout the entire software development life cycle, which provides the essential value of what the software is and what it does with data from the user's viewpoint.

Santillo attempts to use the Function Point method to measure software size in an SOA environment [7]. After comparing the effect of adopting the first and second-generation methods (IFPUG and COSMIC respectively), Santillo identifies several critical issues. The prominent one is that SOA is functionally different from traditional software architectures, because the "function" of a service should represent a real-world self-contained business activity [8].

More issues appear when applying IFPUG to software system size measurement. For example, the effort of wrapping legacy code and data to work as services cannot be assigned to any functional size. Measuring with the COSMIC approach, on the contrary, is supposed to satisfy the typical sizing aspects of SOA-based software. However, there is a lack of guidelines for practical application of COSMIC measurement in SOA context.

In addition to the application of Function Points, Liu et al. [9] use Service Points to measure the size of SOA-based software. The software size estimation is based on the sum of the sizes of each service.

$$Size = \sum_{i=1}^{n}(P_i \times P) \qquad (1)$$

Where $P_i$ is an infrastructure factor with empirical value that is related to the supporting infrastructure, technology and governance processes. $P$ represents a single specific service's estimated size that varies with different service types, including existing service, service built from existing resources, and service built from scratch. This approach implies that the size of a service-oriented application depends significantly on the service type. However, the calculation of $P$ for various services is not discussed in detail.

### C. SMAT-AUS Framework

SMAT-AUS [10] is an ongoing framework that is developed to determine the scope and estimate cost and effort for SOA projects. This framework reveals not only technical dimension but also social, cultural, and organizational dimensions of SOA implementation. When applying the SMAT-AUS framework to SOA-based software development, Service Mining, Service Development, Service Integration and SOA Application Development are classified as separate SOA project types. For each SOA project type, a set of methods, templates and cost models and functions are used to support the cost and effort estimation work for each project time which are then used to generate the overall cost of an SOA project (a combination of one or more of the project types).

Except for the SMART (Software Engineering Institute's Service Migration and Reuse Technique) method [11] that can be adopted for service mining cost estimation, currently there are no other metrics suitable for the different projects beneath the SMAT-AUS framework. Instead, some abstract cost-estimation-discussions related to aforementioned project types can be found through a literature review. Umar and Zordan [12] warn that both gradual and sudden migration would be expensive and risky so that costs and benefits must be carefully weighed. Bosworth [13] gives a full consideration about complexity and cost when developing Web services. Liu et al. [9] directly suggest that traditional methods can be used to estimate the cost of building services from scratch. Since utilizing solutions based on inter-operable services is part of service-oriented integration (SOI) and results in an SOI structure, Erl [3] gives a bottom line of effort and cost estimation for cross-application integration:

"The cost and effort of cross-application integration is significantly lowered when applications being integrated are SOA-compliant."

A generic SOA application could be sophisticated and comprise a combination of project types listed above. This is handled in SMAT-AUS by breaking the problem into more manageable pieces (i.e. a combination of project types) however specifying how all of these pieces are estimated and the procedure required for practical estimation of software development cost for SOA-based systems is still being developed.

*D. Other Approaches*

Discussion about cost estimation for SOA implementation also appears in industry. Linthicum [14] outlines some general guidelines for estimating the cost of an SOA application. According to these guidelines, the calculation of SOA cost can be expressed as a sum of several cost analysis procedures.

$$\begin{aligned}Cost\ of\ SOA = ( & Cost\ of\ Data\ Complexity \\ & + Cost\ of\ Service\ Complexity \\ & + Cost\ of\ Process\ Complexity \\ & + Enabling\ Technology\ Solution)\end{aligned} \quad (2)$$

Furthermore, Linthicum also provides some detailed specification. For example, the basic element Complexity of the Data Storage Technology is figured as a percentage between 0% and 100% (Relational is 30%, Object-Oriented is 60%, and ISAM is 80%). Nevertheless, the other aspects of the calculation are suggested to follow similar means without clarifying essential matters. Meanwhile, Linthicum reminds that the notable problem is that this approach is not a real metric. Additionally, SOA based software is inevitably more complicated than traditional software [8]. It is therefore doubtful that Data Complexity, System Complexity, Service Complexity and Process Complexity are sufficient to represent the complexity of SOA-based systems.

As shown, both academia and industry have published little work relating to estimating costs for SOA-based software. In particular, there is not a solution to satisfy the development cost estimation for SOA-based software. We attempt to address these issues by providing a SOA cost estimation framework in this paper.

III. SOA-BASED SOFTWARE COST ESTIMATION USING DIVIDE-AND-CONQUER APPROACH

*A. Divide-and-Conquer (D&C)*

The history of D&C method can be traced back to as early as 200BC [19], when the Babylonian reciprocal table of Inakibit-Anu was used to facilitate searching and sorting numerical values. However, the first description of the D&C algorithm appears in John Mauchly's article discussing its application in computer sorting [19]. Nowadays, the D&C approach is applied widely in areas such as Parallel Computing [20], Clustering Computing [21], Granular Computing [22], and Huge Data Mining [23].

The principle underlying D&C is shown in Figure 1. That is to recursively <u>decompose</u> the problem into smaller sub-problems until all the sub-problems are sufficiently simple enough, and then to <u>solve</u> the sub-problems. Resulting solutions are then <u>recomposed</u> to form an overall solution.

Adopting the principle, the D&C procedure will lead to different subroutines for different sub-problems. Normally, some or all of the sub-problems are of the same type as the input problem, thus D&C procedure can be naturally expressed recursively. The QuickSort [19] algorithm is also such a procedure.

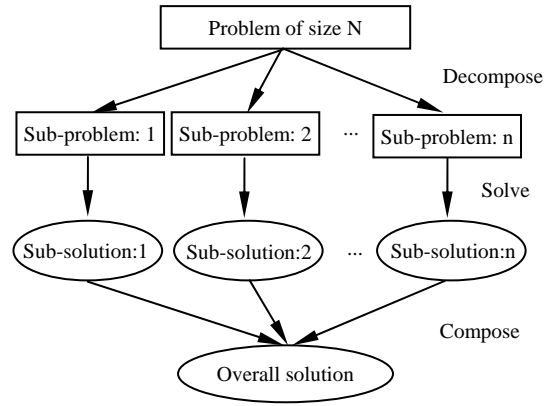

Figure 1. Principle of Divide-and-Conquer.

The advantages of applying the D&C approach to suitable problems are multifold, and can be classified as the following:

- *Structural Simplicity*. Profiting from perhaps the simplest structuring technique, D&C has been identified as a high prior strategy to resolve problems not only in the computer science field but also in politics and sociology fields. No mater where the D&C approach is applied the solution structure can be expressed explicitly in a program-like function such as:

$$\begin{aligned}Solution(x) \equiv \\ &IF\ IsBase(x) \\ &THEN\ SolveDirectly(x) \\ &ELSE\ Compose(Solution(Decompose(x)))\end{aligned} \quad (3)$$

Where $x$ is the original problem that will be solved through *Solution* procedure. *IsBase* is used to verify whether the problem $x$ is primitive or not, which returns TRUE if $x$ is a basic problem unit, or FALSE otherwise. *SolveDirectly* presents the conquer procedure. *Decompose* is referred to as the decomposing operation, while *Compose* is referred to as the composing operation.

- *Computational Efficiency.* D&C is frequently used for designing fast algorithms. In appropriate application scenarios, the D&C approach leads to asymptotically optimal cost for solving the problems. Assure a problem of size $N$ can be broken into a bounded number $P$ of sub-problems of size $N/P$ step by step, and all the basic sub-problems have constant-bounded size. Then the D&C algorithm will have O($N\log N$) worst-case program execution performance. Normally, The consequence is more flexible because the size and the number of tasks can be decided at run-time.
- *Parallelism.* Since sub-problems in the individual division stage are logically and physically independent, the D&C approach can be naturally executed in parallel procedures. For computing problems, D&C is suitable for application in parallel machines due to not only the independent problem grains but also the efficient use of cache and deep memory hierarchies [24]. In fact, D&C has been considered as one of the well-known parallel programming paradigms.
- *Capability of Solving Complexity.* Through breakdown of an overall goal into smaller and independent sub-problems, the D&C strategy provides adaptation scalability and variability, and is frequently used in the areas of engineering to reduce and manage complexity. Those complicated cases, such as resolutions for conceptually difficult problems, and approximate algorithms for NP-hard problems, are usually based on the D&C principle.

Given these merits, D&C can be considered a suitable and effective approach to accommodate complex problems such as cost estimation for SOA-based software development, where individual measures must be carried out independently. The following sections discuss its applications in SOA cost estimation.

### B. Service Classification

Implementing SOA could be complex and onerous, while complexity measurement for SOA-based system is still an open question [15]. Chaos [16] even claims that the complexity is restricting some SOA implementations. For the same reason, there are also many challenges to estimate the cost and effort of SOA-based software development.

Fortunately, the advantages of SOA are mainly reusability and composability with an emphasis on extensibility and flexibility, at a high level of granularity and abstraction. In other words, SOA-based software can be naturally divided into a set of loosely coupled services. These services can then be classified through their different features. Krafzig et al. [17] has identified that distinguishing services into classes is extremely helpful when properly estimating the implementation and maintenance cost, and the cost factors may vary depending on the service type. However, there is not a standard way to categorize services. Service classification can be different for different purposes, for example differentiating services according to their target audience [8], categorizing services through their business roles and responsibilities [3], and classifying services by using their background techniques and protocols [18]. As we focus on the development process, services in our work are distinguished as follows:

- *Available Service* (basic service type), is the service already existing i.e. is it provided by a third party or inherited from legacy SOA based systems.
- *Migrated Service* (basic service type), is the service to be generated through modifying or wrapping reusable traditional software component(s).
- *New Service* (basic service type), is the service to be developed from scratch.
- *Combined Service* is the service any combination of above three types of basic services or other combined services.

Through this type of classification, four different development areas are identified in SOA projects. These areas present both a decomposition process that results in Service Discovery, Service Migration, and Service Development, and a recomposition process that is Service Integration. The cost estimation for overall SOA-based software development can then be separated into these smaller areas with corresponding metrics. Therefore, the D&C approach is a feasible attempt for SOA-based software cost estimation following this development oriented service classification.

### C. D&C Cost Estimation Framework

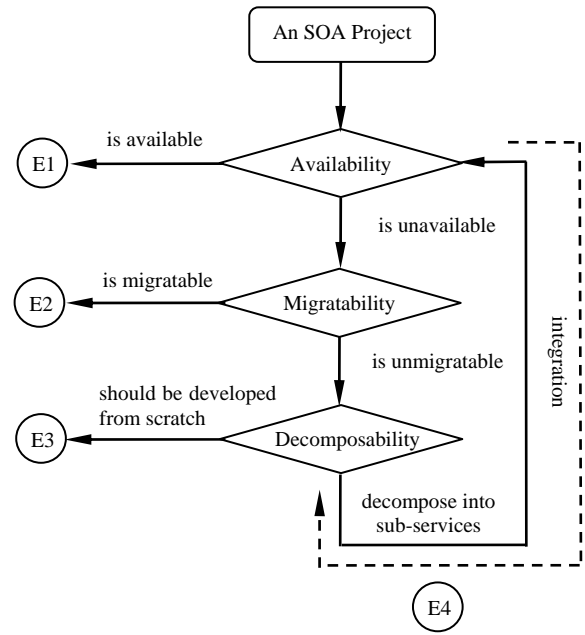

Figure 2. Procedure of SOA Project Development Cost Estimation based on Divide-and-Conquer.

The notion of the proposed cost estimation framework for SOA-based software closely follows the D&C principle. Firstly, through the service-oriented analysis, the SOA project is divided into basic services recursively. Secondly, different sets of metrics are adopted to satisfy the cost and effort estimation for different service development processes. The total cost and effort of the SOA project will be calculated through the service integration procedure as shown in Figure 2.

Where E1 is the cost estimation model or software size measurement used to accomplish modeling or sizing work for discovering available services, E2 represents migrating potential services, E3 represents developing new services, and E4 is the cost estimation model or size measurement for calculating the service integration effort. As for the Decomposability condition, particularly, it depends on the design and real situations whether the current service should be further divided or developed as a whole. The framework in Figure 2 presents the generic process of SOA cost estimation using the D&C method. The process can lead to both a model tree if applying D&C to modeling the development cost estimation for SOA-based software, and a sizing tree if applying D&C to measuring the size of an SOA-based application. To calculate the ultimate cost and effort, the predicted size should be combined as a parameter with the estimation model.

To precisely describe the D&C based cost estimation for SOA-based software development, the complete process can be expressed in the following pseudo code. Here we define the stage where that service division occurs as the service levels, and the combined service stands in a higher level next to its successive component services.

TABLE I. ALGORITHM OF SOA PROJECT DEVELOPMENT COST ESTIMATION BASED ON DIVIDE-AND-CONQUER

```
1)//Treat the project as the highest-level service S to analyze.
2)double SoaCostEstimation(service S) {
3)    double cost = 0;
4)    Determine the type of S according to the design and real
      situations.
5)    switch (the type of S) {
6)       case AVAILABLE:
7)          cost += The cost of service discovery;
8)          break;
9)       case MIGRATABLE:
10)         cost += The cost of service migration (service wrapping);
11)         break;
12)      case NEW:
13)         cost += The cost of service development;
14)         break;
15)      default:
16)         Divide S into component services at lower level.
17)         foreach component service in S
18)            cost += SoaCostEstimation(component service);
19)         cost += The cost of service integration for component
                    services in S;
20)         break;
21)   }
22)   return cost;
23) }
```

As shown in Table 1, the SOA project itself is treated as the highest-level coarse-grain service, which is also the initial input parameter of SoaCostEstimation function. Within the body of SoaCostEstimation function, the cost of the input service development will be estimated directly if the service belongs to those three basic types, or recursively calculated by analyzing and composing the cost and effort of the development for component services. When composing individual service development costs into the overall SOA-based software development cost, the strategy of supposed service integration is progressed level-by-level instead of integrating the services all at once. The reason of adopting such a strategy is that, according to our work, service integration occurring in different levels will make different contributions to the total cost and effort of the project development.

A real example can be used to show the application process of the D&C based cost estimation framework for SOA-based software development in practice, which is demonstrated in the next section.

IV. AN APPLICATION CASE STUDY

We employ the RailCo Ltd. case study presented in [3]. There are two reasons for choosing this case: The RailCo Ltd. case study characterizes all the service types listed in the previous section, and there are a limited number of services, which are adequate for illustrative purposes in this paper.

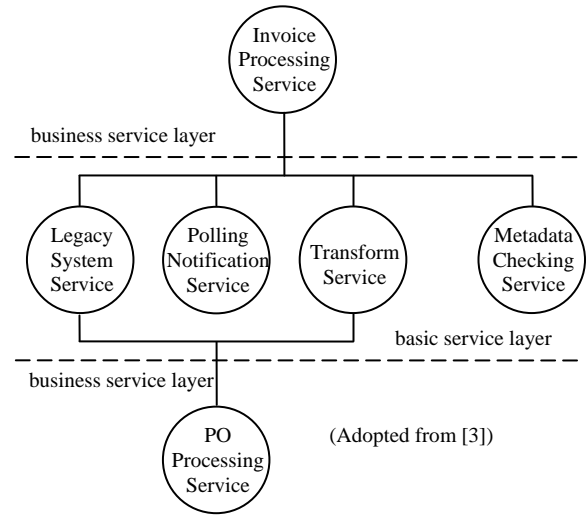

Figure 3. Redesigned Automation System of RailCo Ltd.

RailCo Ltd. is a railway parts supplier company specializing in air brakes and related installation tools. To improve the working efficiency of this company, a service-oriented analysis was conducted, which decomposed the business process logic into a series of service candidates. RailCo Ltd. revealed the requirements of two business services in higher level and four application services in lower level. The redesigned automation system is represented in Figure 3 following current disciplines:

*1)* Legacy System Service is migrated from the previous project.

*2)* Polling Notification Service and Transform Service are new services that should be developed from scratch.

*3)* Metadata Checking Service is an available service provided by a third party.

*4)* Invoice Processing Service and PO Processing Service are both combined services containing all or some of above basic services.

The procedure of cost and effort estimation for developing this redesigned service-oriented project is illustrated in Figure 4. The detailed steps are elaborated as following:

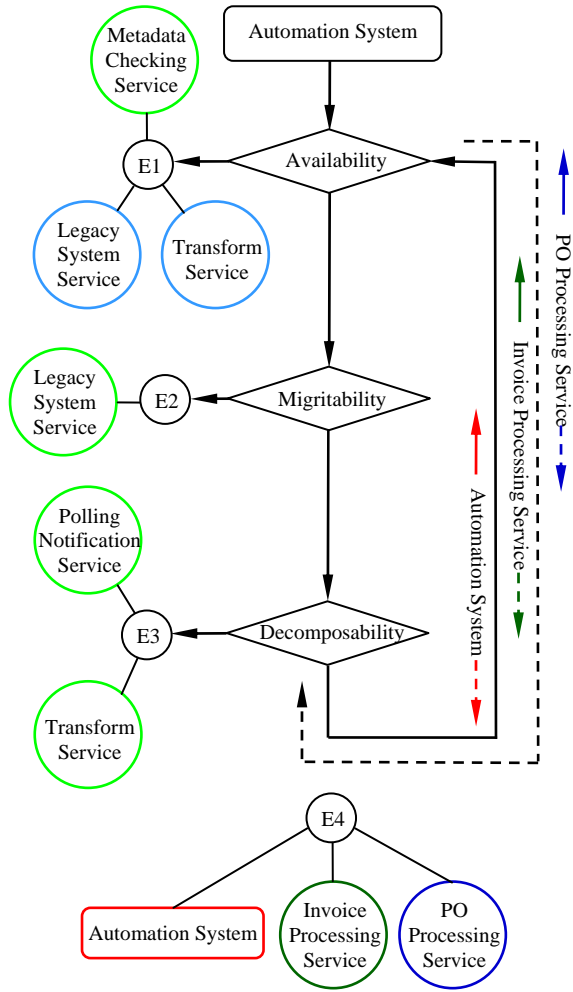

Figure 4. Procedure of Cost and Effort Estimation for Developing an Automated System (RailCo Ltd. case study).

*1)* Divide the Automation System into an Invoice Processing Service and a PO Processing Service.

*2)* Divide the Invoice Processing Service into its four basic component services.

*3)* Estimate the cost and effort of discovering the available Metadata Checking Service by using corresponding metrics E1.

*4)* Estimate the cost and effort of migrating the Legacy System Service by using corresponding metrics E2.

*5)* Estimate the cost and effort of developing the Polling Notification Service and Transform Service by using corresponding metrics E3.

*6)* Estimate the cost and effort of integrating the above four component services into the Invoice Processing Service by using corresponding metrics E4.

*7)* Divide the PO Processing Service into its two basic component services.

*8)* Notice that Legacy System Service and Transform Service have both been taken into account.

*9)* Estimate the cost and effort of mining the Legacy System Service and Transform Service by using corresponding metrics E1. Since these two services are in the same project and can be directly identified, the cost and effort here can be treated as zero in this special case.

*10)* Estimate the cost and effort of integrating the above two component services into the PO Processing Service by using the corresponding metrics E4.

*11)* Estimate the cost and effort of integrating the Invoice Processing Service and PO Processing Service into the Automation System by using the corresponding metrics E4.

*12)* Sum up all the estimation results to calculate the total cost and effort of the Automation System development.

Through the demonstration of the RailCo Ltd. case, the D&C framework is proven helpful for simplifying and regulating the SOA-based software cost estimation. Moreover, all the simplified cost estimation problems are independent enough to be solved in parallel. The uniform and explicit working procedure within this D&C framework is then a feasible attempt to SOA based software cost estimation.

V. DISCUSSION

Unlike many of the current cost estimation approaches mentioned in section II, the proposed D&C framework uses a set of metrics to satisfy the development cost estimation for SOA-based software. A similar strategy is also adopted in the SMAT-AUS framework. When comparing with the SMAT-AUS framework, the D&C framework is distinct because:

- Regarding SOA based software cost estimation, the SMAT-AUS framework tries to cover the full scope of factors and the whole project life cycle, whilst this D&C framework concentrates on the software development procedure.
- Considering the effort of selecting available services, this D&C framework lists Service Discovery as an individual cost estimation area as well as Service Migration, Service Development, and Service Integration.
- The SMAT-AUS framework lists the development cost estimation for complete SOA application and for services as equal SOA projects. The D&C framework instead estimates the overall cost and effort through the independent estimation activities in four different development areas of an SOA application.
- The D&C framework is generic and flexible. Through switching different type of metrics, this proposed framework could satisfy different requirements of SOA-based software cost estimation. Such as building cost estimation model, measuring software size, and predicting the overall cost ultimately.

One issue is that there are currently few available metrics for the detailed cost estimation for SOA-based software development. Future research should develop new metrics to resolve this issue. Meanwhile, some reusable existing metrics can be integrated into the proposed D&C framework, for example Tansey and Stroulia's work [6] related to Service Development and SMART method [11] related to Service Migration. Over all, instead of trying to enumerate SOA project types, the D&C framework unifies and regulates the cost and effort estimation for SOA-based software development.

## VI. CONCLUSION AND FUTURE RESEARCH

Software cost estimation plays a vital role in software development projects, especially for SOA-based software development. However, current cost estimation approaches for SOA-based software are inadequate due to the architectural difference and the complexity of SOA applications. This paper proposes a D&C cost estimation framework for SOA-based software development. Based on the principle of D&C theory, this framework can be helpful for simplifying the complexity of SOA cost estimation. By hosting different sets of metrics, this generic framework will be suitable not only for the complete cost estimation work but also for the partial requirements, such as building estimation model, and measuring the size of SOA applications.

Given the requirement of developing these metrics, future research should be carried out in multiple areas:

- *Modeling and sizing the cost and effort of service discovery.* There are challenges of performance bottleneck problems when dealing with service selection, because a large number of Web service descriptions could be published in UDDI (Universal Description, Discovery and Integration) registries. Cost and effort varies with different service discovery techniques, such as semantic annotation, quality of service, and centralized registry implementations in respective enterprise communication groups.
- *Modeling and sizing the cost and effort of service migration, including service wrapping.* Wrapping and migrating legacy and existing system components are normally appropriate approaches to reuse resources maximally within the SOA environment. However, service migration could be an expensive and risky undertaking with costs and benefits that must be weighted carefully. Black-box, grey-box, and white-box migration approach types should be taken into account respectively. Available service-oriented migration techniques, for example SMART [11], can be adopted to support this work area.
- *Modeling and sizing the cost and effort of service development.* Since a service should support messaging-based communication and dynamic hardware scaling, the development cost must be estimated according to developing the infrastructure, creating a scalable data centre, and deploying required hardware. In addition, as services always present self-contained business activities, the business factor should also be considered seriously in service development. Traditional software cost estimation approaches can be reused when developing services traditionally.
- *Modeling and sizing the cost and effort of service integration.* There is strong motivation for assembling services into larger scale service or composite applications in modern enterprises with an SOA infrastructure. These include increasing stability, decreasing development cost, and quickening business response. Cost estimation metrics for service integration should be developed with consideration of different integration strategies, for example point-to-point integration or full integration by using ESB (Enterprise Service Bus).

Since the importance of service integration is being realized at different speeds in industry, our current work is mainly concentrating on cost estimation for Service Integration.

ACKNOWLEDGMENT

NICTA is funded by the Australian Government as represented by the Department of Broadband, Communications and the Digital Economy and the Australian Research Council through the ICT Centre of Excellence program.

REFERENCES

[1] Z. Stojanovic and A. Dahanayake, Service-Oriented Software System Engineering: Challenges and Practices. Hershey, PA: IGI Global, Apr. 2005.

[2] E. Jamil, "SOA in Asynchronous Many-to-One Heterogeneous Bi-Directional Data Synchronization for Mission Critical Applications," WeDoWebSphere, Jul. 2009. [Online]. Available: http://wedowebsphere.de/news/1528/-SOA%20in%20Asynchronous%20Many-to-one%20Heterogeneous%20Bi-Directional%20Data%20Synchronization%20. [Accessed: Nov. 2009].

[3] T. Erl, Service-Oriented Architecture: Concepts, Technology, and Design. Crawfordsville: Prentice Hall PTR, Aug. 2005.

[4] B.W. Boehm, C. Abts, A.W. Brown, S. Chulani, B.K. Clark, E. Horowitz, R. Madachy, D.J. Reifer, and B. Steece, Software Cost Estimation with COCOMO II. New Jersey: Prentice Hall PTR, Aug. 2000.

[5] I. Sommerville, Software Engineering, 8th ed.. London: Addison Wesley, Jun. 2006.

[6] B. Tansey and E. Stroulia, "Valuating Software Service Development: Integrating COCOMO II and Real Options Theory," Proc. the First International Workshop on the Economics of Software and Computation, IEEE Press, May 2007, pp. 8-8, doi: 10.1109/ESC.2007.11.

[7] L. Santillo, "Seizing and Sizing SOA Application with COSMIC Function Points," Proc. the 4th Software Measurement European Forum, Rome, Italy, May 2007.

[8] N. M. Josuttis, SOA in Practice: The Art of Distributed System Design, Sebastopol: O'Reilly Media, Inc., 2007.

[9] J. Liu, Z. Xu, J. Qiao, and S. Lin, "A Defect Prediction Model for Software Based on Service Oriented Architecture using EXPERT COCOMO," Proc. Chinese Control and Decision Conference (CCDC '09), IEEE Press, Jun. 2009, pp. 2591-2594, doi: 10.1109/CCDC.2009.5191800.

[10] L. O'Brien, "A Framework for Scope, Cost and Effort Estimation for Service Oriented Architecture (SOA) Projects," Proc. 20th Australian Software Engineering Conference (ASWEC'09), IEEE Press, Apr. 2009, pp. 101-110, doi: 10.1109/ASWEC.2009.35.

[11] G. Lewis, E. Morris, L. O'Brien, D. Smith and L. Wrage, "SMART: The Service-Oriented Migration and Reuse Technique," CMU/SEI-2005-TN-029, Software Engineering Institute, USA, Sept. 2005.

[12] A. Umar and A. Zordan, "Reengineering for Service Oriented Architectures: A Strategic Decision Model for Integration versus Migration," Journal of Systems and Software, vol. 82, Mar. 2009, pp. 448-462, doi: 10.1016/j.jss.2008.07.047.

[13] A. Bosworth, "Developing Web Services," Proc. 17th International Conference on Data Engineering (ICDE 2001), IEEE Press, Apr. 2001, pp. 477-481, doi: 10.1109/ICDE.2001.914861.

[14] D. Linthicum, "How Much Will Your SOA Cost?," SOAInstitute.org, Mar. 2007. [Online]. Available: http://www.soainstitute.org/articles/article/article/how-much-will-your-soa-cost.html. [Accessed: Nov. 2009].

[15] D. Norfolk, "SOA Innovation and Metrics," IT-Director.com, Dec. 2007. [Online]. Available: http://www.it-director.com/business/change/content.php?cid=10146. [Accessed: Nov. 2009].

[16] D. Chaos, "SOA is not dead, but complexity is killing some implementations," Technoracle (a.k.a. "Duane's World"), Jan. 2009. [Online]. Available: http://technoracle.blogspot.com/2009/01/soa-is-not-dead-but-complexity-is.html. [Accessed: Jul. 2009].

[17] D. Krafzig, K. Banke, and D. Slama, Enterprise SOA: Service-Oriented Architecture Best Practices, Upper Saddle River: Prentice Hall PTR, Nov. 2004.

[18] J. Davies, D. Schorow, S. Ray, and D. Rieber, The Definitive Guide to SOA: Oracle Service Bus, 2nd ed.. New York: Apress, Sept. 2008.

[19] D. E. Knuth, The Art of Computer Programming: Volume 3, Sorting and Searching, 2nd ed.. Reading, MA: Addison-Wesley Professional, May 1998.

[20] Y. Bai and R. C. Ward, "A Parallel Symmetric Block-Tridiagonal Divide-and-Conquer Algorithm," Transactions on Mathematical Software (TOMS), vol. 33, Aug. 2007, pp. A25, doi: 10.1145/1268776.1268780.

[21] M. Khalilian, F. Z. Boroujeni, N. Mustapha, and M. N. Sulaiman, "K-Means Divide and Conquer Clustering," Proc. the 2nd International Conference on Computer and Automation Engineering (ICCAE 2009), IEEE Press, Mar. 2009, pp. 306-309, doi: 10.1109/ICCAE.2009.59.

[22] T. Y. Lin, "Divide and Conquer in Granular Computing Topological Partitions," Proc. Annual Meeting of the North American Fuzzy Information Processing Society (NAFIPS 2009), IEEE Press, Jun. 2005, pp. 282-285, doi: 10.1109/NAFIPS.2005.1548548.

[23] F. Hu and G. Wang, "Huge Data Mining Based on Rough Set Theory and Granular Computing," Proc. IEEE/WIC/ACM International Conference on Web Intelligence and Intelligent Agent Technology (WI-IAT'08), IEEE Press, Dec. 2008, pp. 655-658, doi: 10.1109/WIIAT.2008.84.

[24] C. Zhang and B. Xue, "Divide-and-Conquer: A Bubble Replacement for Low Level Caches," Proc. the 23rd International Conference on Supercomputing (ICS'09), ACM, Jun. 2009, pp. 80-89, doi: 10.1145/1542275.1542291.

[25] J. Cardoso, "How to Measure the Control-Flow Complexity of Web Processes and Workflows," Workflow Handbook 2005, Layna Fischer, Apr. 2005, pp. 199-212.